\documentclass[lettersize,journal]{IEEEtran}

\usepackage{amsmath,amssymb,bm}
\usepackage{graphicx}
\usepackage{cite}
\usepackage{url}
\usepackage{subcaption}
\usepackage{flushend}

\newcommand{\E}{\mathbb{E}}
\newcommand{\Prb}{\Pr}
\newcommand{\dd}{\mathrm{d}}
\newcommand{\JACD}{\Xi_{\rm JC}}
\newcommand{\JAFD}{\Gamma_{\rm JF}}
\newcommand{\Rvec}{\bm R}
\newcommand{\Vvec}{\dot{\bm R}}
\newcommand{\pos}[1]{\left[#1\right]^+}

\begin{document}

\title{Joint Average Contiguous Duration of Correlated Fading Envelopes}

\author{Muhammad Adil, Syed Junaid Nawaz, Ernestina Cianca, Wali Ullah Khan, and Salman A. AlQahtani
\thanks{Muhammad Adil and Ernestina Cianca are with the Department of Electronics Engineering, University of Rome Tor Vergata, 00133 Rome, Italy (e-mail: muhammad.adil@uniroma2.it; cianca@ing.uniroma2.it).}
\thanks{Syed Junaid Nawaz is with the Department of Computer Engineering, COMSATS University Islamabad, Islamabad, Pakistan (e-mail: junaidnawaz@gmail.com).}
\thanks{Wali Ullah Khan is with the Interdisciplinary Centre for Security, Reliability and Trust (SnT), University of Luxembourg, Luxembourg (e-mail: waliullahkhan30@gmail.com).}
% \thanks{Salman A. AlQahtani is with the New Emerging Technologies and 5G Network and Beyond Research Chair, Department of Computer Engineering, College of Computer and Information Sciences, King Saud University, Riyadh, Saudi Arabia (e-mail: salmanq@ksu.edu.sa).}
% \thanks{This work was supported by the Ongoing Research Funding Program—Research Chairs, King Saud University, Riyadh, Saudi Arabia, under Grant ORF-RC-2026-5300.}
\thanks{Salman A. AlQahtani is with the New Emerging Technologies and 5G Network and Beyond Research Chair, King Saud University, Riyadh, Saudi Arabia (e-mail: salmanq@ksu.edu.sa).}
\thanks{This work was supported by the Ongoing Research Funding Program-Research Chairs, King Saud University, Riyadh, Saudi Arabia, under Grant ORF-RC-2026-5300.}

}

\markboth{}{}

\maketitle

\begin{abstract}
Average fade and average contiguous durations measure how long an envelope stays continuously below one threshold or between two. Neither captures how long correlated envelopes stay simultaneously within their intervals, which governs usable runs in physical-layer secret-key generation (SKG). We characterize this joint persistence by the joint average contiguous duration (JACD), whose entry rate depends on joint boundary crossings, so marginal statistics generally cannot determine it. We derive closed-form JACD and joint average fade duration expressions for correlated Rayleigh envelopes. In reciprocal SKG, JACD-balanced quantization equalizes usable run lengths and lowers bit disagreement and information disclosed to the eavesdropper.
\end{abstract}


\section{Introduction}
\IEEEPARstart{S}{econd-order} fading statistics describe how a channel envelope behaves over time. Two classical metrics of this kind, the level-crossing rate (LCR) and average fade duration (AFD), follow from Rice's crossing analysis~\cite{rice1945noise}. For a single threshold, the AFD is the mean duration of fade-duration (FD) instances, each lasting from a downward crossing of the threshold to the next upward crossing. For an amplitude range of interest (ARoI) bounded by two thresholds, the average contiguous duration (ACD)~\cite{9431101} is the mean duration of contiguous-duration (CD) instances, each lasting from an entry into the ARoI to the next exit. Entry and exit can each occur through either threshold. The ACD reduces to the AFD when the lower threshold is zero and to the average non-fade duration when the upper threshold tends to infinity. Since every bin of a multilevel quantizer can be viewed as an ARoI, the ACD is directly relevant to multilevel quantization. The ACD coincides with the average stay duration defined earlier in~\cite{abdi2003lcr}; \cite{9431101} derived closed-form ACD expressions for several fading models, and \cite{9500118} applied the ACD to multilevel quantization for physical-layer secret-key generation (SKG).

The ACD of a single envelope is insufficient when two channel observations must remain usable simultaneously. This is the case in reciprocal SKG, where Alice and Bob quantize correlated measurements of the same propagation channel~\cite{9500118}. A usable run begins once both observations are inside their prescribed quantization intervals and ends when either one leaves. We call such a run a joint contiguous-duration (JCD) instance and its mean the joint average contiguous duration (JACD). The JACD predicts the average number of consecutive samples that both nodes can use, which can guide the choice of probing rate and quantization intervals. Similarly, a joint fade-duration (JFD) instance lasts while both envelopes are below their fade thresholds, and its mean is the joint average fade duration (JAFD).

The JACD equals the joint occupancy probability divided by the joint entry rate. An entry occurs when one envelope crosses into its interval while the other is already inside, so the entry rate depends on how the envelopes behave jointly at the boundaries. Hence, the JACD generally cannot be determined from the marginal ACDs or AFDs. Reciprocal SKG lies between the independent and identical-envelope cases, because Alice's and Bob's observations are strongly but not perfectly correlated. Correlated crossing and fade-duration statistics have been studied for diversity systems~\cite{adachi1988effects,1175224}, and bivariate Rayleigh envelope models are well established~\cite{846501}. The diversity studies, however, analyze a single combiner output against one threshold; their equal-threshold joint fade gives a consistency check, since the JAFD then reduces to the dual-branch selection-combining result~\cite{adachi1988effects}. Abdi et al.~\cite{abdi2003lcr} also defined a vector average stay duration inside a hypercube around a reference channel realization, which measures how fast the channel changes. The JACD instead concerns how long correlated envelopes stay together within fixed, generally unequal, amplitude intervals, as in quantization. To the best of our knowledge, this joint persistence has not been characterized.

The contributions are threefold. First, building on the multidimensional crossing formula of~\cite{belyaev1968exits} and the vector average stay duration of~\cite{abdi2003lcr}, we formulate the joint persistence of $K$ correlated envelopes within fixed amplitude intervals and express the entry rate of the resulting hyperrectangle through marginal LCRs and conditional interval probabilities. Second, we obtain JAFD as the special case with zero lower thresholds and derive closed-form expressions of both metrics for correlated Rayleigh envelopes under the Jakes model. These expressions allow unequal intervals, average powers, and Doppler spreads, and recover the independent limit and, for equal Doppler spreads, the perfectly correlated limit. Third, we evaluate JACD-balanced multilevel quantization in a reciprocal SKG chain with blind polar-code reconciliation. The results show that this design makes usable run lengths more uniform across bins and reduces bit disagreement and public disclosure.

\section{JACD Formulation}

Let $\Rvec(t)\triangleq[R_1(t),\ldots,R_K(t)]^{\mathsf T}$ be a stationary, ergodic, almost surely continuously differentiable $K$-dimensional envelope process, where $K\ge1$, $(\cdot)^{\mathsf T}$ denotes transpose, and let $\Vvec(t)\triangleq\dd\Rvec(t)/\dd t$. For an admissible region $\mathcal D\subset\mathbb R^K$ whose boundary $\partial\mathcal D$ is piecewise continuously differentiable ($C^1$), a joint contiguous-duration (JCD) instance is a maximal interval $(t_{\rm in},t_{\rm out})$ over which $\Rvec(t)\in\mathcal D$. Its duration is $T_{\mathcal D}\triangleq t_{\rm out}-t_{\rm in}$ and its mean over JCD instances is $\Xi_{\mathcal D}\triangleq\E[T_{\mathcal D}]$, where $\E[\cdot]$ denotes expectation. The stationary occupancy probability is $P_{\mathcal D}\triangleq\Prb\{\Rvec\in\mathcal D\}=\int_{\mathcal D}f_{\Rvec}(\bm r)\,\dd\bm r$, where $\Prb\{\cdot\}$ denotes probability, $\bm r$ is a realization of $\Rvec$, and $f_{\Rvec}$ is the joint envelope probability density function (PDF).

For a regular boundary point $\bm r\in\partial\mathcal D$, let $\bm n(\bm r)$ be the outward unit normal and let $\bm v$ denote a realization of $\Vvec$. By the multidimensional Rice crossing formula~\cite{belyaev1968exits}, the total inward boundary-crossing intensity is
\begin{equation}
\Lambda_{\mathcal D}^{\rm in}
=\int_{\partial\mathcal D}\int_{\mathbb R^K}
\pos{-\bm v^{\mathsf T}\bm n(\bm r)}
 f_{\Rvec,\Vvec}(\bm r,\bm v)
\,\dd\bm v\,\dd S(\bm r),
\label{eq:general_flux}
\end{equation}
where $[x]^+\triangleq\max(x,0)$, $f_{\Rvec,\Vvec}$ denotes the joint PDF of $\Rvec$ and $\Vvec$, and $\dd S$ is surface measure. Assuming finite nonzero occupancy and inward flux, with tangential and simultaneous multiple-face contacts having zero intensity, stationarity gives
\begin{equation}
\Xi_{\mathcal D}=\frac{P_{\mathcal D}}{\Lambda_{\mathcal D}^{\rm in}}.
\label{eq:master}
\end{equation}
Thus, the mean joint-excursion duration equals the fraction of time spent in the admissible region divided by the rate at which new excursions enter it, as in the vector average stay duration of~\cite{abdi2003lcr}.

For the hyperrectangle $\mathcal D\triangleq\prod_{k=1}^{K}(a_k,b_k]$, with $0\le a_k<b_k$, define $\mathcal D_{-k}\triangleq\prod_{j\ne k}(a_j,b_j]$ and let $\bm r_{-k}$ collect all envelope components except the $k$th. Let $\dot R_k\triangleq\dd R_k/\dd t$, and let $f_{\Rvec,\dot R_k}$ denote the joint PDF of $\Rvec$ and $\dot R_k$. The lower- and upper-face inward rates are
\begin{align}
\Lambda_{k,a_k}
&=\int_{\mathcal D_{-k}}\int_0^{\infty}
 v f_{\Rvec,\dot R_k}(\bm r_{-k},a_k,v)
\,\dd v\,\dd\bm r_{-k},
\label{eq:face_lower}\\
\Lambda_{k,b_k}
&=\int_{\mathcal D_{-k}}\int_{-\infty}^{0}
 (-v) f_{\Rvec,\dot R_k}(\bm r_{-k},b_k,v)
\,\dd v\,\dd\bm r_{-k}.
\label{eq:face_upper}
\end{align}
Hence $\Lambda_{\mathcal D}^{\rm in}=\sum_{k=1}^{K}(\Lambda_{k,a_k}+\Lambda_{k,b_k})$. If each $\dot R_k$ is independent of the instantaneous envelope vector, let $N_k(r)$ denote the one-sided marginal LCR and define $q_{-k|k}(r)\triangleq\Prb\{R_j\in(a_j,b_j],\,\forall j\ne k\mid R_k=r\}$. Then
\begin{equation}
\Lambda_{\mathcal D}^{\rm in}
=\sum_{k=1}^{K}\!\left[
N_k(a_k)q_{-k|k}(a_k)+N_k(b_k)q_{-k|k}(b_k)
\right],
\label{eq:K_reduced_flux}
\end{equation}
which, together with \eqref{eq:master}, gives the hyperrectangular JACD. This decomposition follows the route used for selection-combining LCRs~\cite{adachi1988effects,1175224}, here applied to both faces of every interval.

For the lower-orthant joint-fade region $\mathcal D_{\rm JF}\triangleq\prod_{k=1}^{K}(0,r_{{\rm f},k}]$, where $r_{{\rm f},k}$ is the fade threshold of the $k$th envelope, let $P_{\rm JF}^{(K)}$ and $\Lambda_{\rm JF}^{(K)}$ denote its occupancy and inward-entry rate. The corresponding JAFD is $\Gamma_{\rm JF}^{(K)}\triangleq P_{\rm JF}^{(K)}/\Lambda_{\rm JF}^{(K)}$.

\subsection{Bivariate JACD}
\begin{figure}[t]
\centering
\includegraphics[width=\columnwidth]{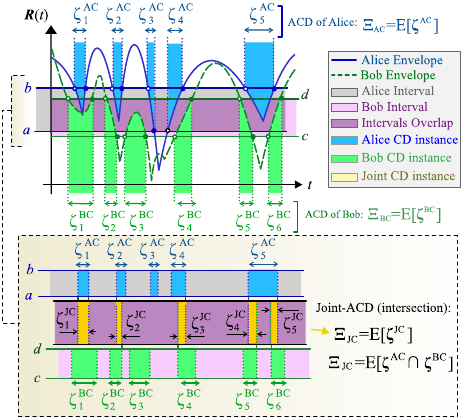}
\caption{JACD formulation from individual CD instances. Highlighted yellow intervals show the maximal temporal intersections of Alice's and Bob's intervals.}
\vspace{-14pt}
\label{ref:fig1}
\end{figure}

Set $K=2$, $R_1=R_{\rm A}$, and $R_2=R_{\rm B}$. Alice and Bob use ARoIs $\mathcal I_{\rm A}\triangleq(a,b]$ and $\mathcal I_{\rm B}\triangleq(c,d]$, respectively, so the joint region is $\mathcal D_{\rm JC}\triangleq\mathcal I_{\rm A}\times\mathcal I_{\rm B}$. A JCD instance is a maximal temporal intersection of Alice's and Bob's scalar CD instances and ends when either envelope leaves its ARoI, see Fig.~\ref{ref:fig1}.

In Fig.~\ref{ref:fig1}, $\xi_i^{\rm AC}$ and $\xi_j^{\rm BC}$ denote the durations of Alice's $i$th and Bob's $j$th scalar CD instances, while $\xi_k^{\rm JC}$ denotes the duration of the $k$th joint intersection. The corresponding ACDs are $\Xi_{\rm AC}\triangleq\E[\xi_i^{\rm AC}]$ for Alice and $\Xi_{\rm BC}\triangleq\E[\xi_j^{\rm BC}]$ for Bob, while $\Xi_{\rm JC}\triangleq\E[\xi_k^{\rm JC}]$ denotes the JACD.

For $K=2$, \eqref{eq:master} becomes
\begin{equation}
\JACD=\frac{P_{\rm JC}}{\Lambda_{\rm JC}},
\label{eq:jacd_general}
\end{equation}
where $x$ and $y$ denote realizations of $R_{\rm A}$ and $R_{\rm B}$, respectively, $P_{\rm JC}\triangleq\int_a^b\int_c^d f_{R_{\rm A},R_{\rm B}}(x,y)\,\dd y\,\dd x$ is the joint-region occupancy probability, and $\Lambda_{\rm JC}$ is its entry rate; from \eqref{eq:K_reduced_flux},
% where $x$ and $y$ denote realizations of $R_{\rm A}$ and $R_{\rm B}$, respectively, $P_{\rm JC}\triangleq\int_a^b\int_c^d f_{R_{\rm A},R_{\rm B}}(x,y),\dd y,\dd x$ is the joint-region occupancy probability, and $\Lambda_{\rm JC}$ is its inward-entry rate; from \eqref{eq:K_reduced_flux},
\begin{align}
\Lambda_{\rm JC}={}&N_{\rm A}(a)q_{\rm B|A}(a;c,d)
+N_{\rm A}(b)q_{\rm B|A}(b;c,d)\nonumber\\
&+N_{\rm B}(c)q_{\rm A|B}(c;a,b)
+N_{\rm B}(d)q_{\rm A|B}(d;a,b),
\label{eq:lambda_sum}
\end{align}
with $q_{\rm B|A}(x;c,d)\triangleq\Prb\{c<R_{\rm B}\le d\mid R_{\rm A}=x\}$ and $q_{\rm A|B}(y;a,b)\triangleq\Prb\{a<R_{\rm A}\le b\mid R_{\rm B}=y\}$. For the joint-fade rectangle $\mathcal D_{\rm JF}=(0,a]\times(0,c]$, let $P_{\rm JF}$ and $\Lambda_{\rm JF}$ denote its occupancy probability and inward-entry rate, respectively; the bivariate JAFD is $\JAFD\triangleq P_{\rm JF}/\Lambda_{\rm JF}$.

\section{Closed-Form Correlated Rayleigh JACD}

Consider Rayleigh envelopes $R_i=|h_i|$, $i\in\{\mathrm A,\mathrm B\}$, where $h_i$ is the complex channel gain and $\Omega_i\triangleq\E[R_i^2]$ is its average power. Let $\rho\in[0,1)$ denote the power-correlation parameter, so the magnitude of the underlying complex correlation coefficient between $h_{\mathrm A}$ and $h_{\mathrm B}$ is $\sqrt{\rho}$; $\rho\to1$ denotes the perfectly correlated limit. Let $I_0(\cdot)$ be the zeroth-order modified Bessel function of the first kind. The joint PDF is \cite{846501}
\begin{align}
f_{R_{\mathrm A},R_{\mathrm B}}(x,y)
&=\frac{4xy}{(1-\rho)\Omega_{\mathrm A}\Omega_{\mathrm B}}
\exp\!\left[-\frac{x^2/\Omega_{\mathrm A}+y^2/\Omega_{\mathrm B}}{1-\rho}\right]
\nonumber\\
&\quad\times I_0\!\left(
\frac{2\sqrt{\rho}\,xy}
{(1-\rho)\sqrt{\Omega_{\mathrm A}\Omega_{\mathrm B}}}
\right),
\qquad x,y\ge0.
\label{eq:biv_rayleigh}
\end{align}

The bivariate Rayleigh cumulative distribution function (CDF) $F_{R_{\mathrm A},R_{\mathrm B}}(x,y) \triangleq\Prb\{R_{\mathrm A}\le x,R_{\mathrm B}\le y\}$ is given in closed form in \cite[Eq.~(6.5)]{simon2004digital}; its
equal-threshold specialization is used in \cite[Eq.~(8)]{6200887}. Define $A_\rho=\sqrt{2/(1-\rho)}$ and $B_\rho=\sqrt{2\rho/(1-\rho)}$, and the normalized amplitudes $\bar{x}=x/\sqrt{\Omega_{\mathrm A}}$ and $\bar{y}=y/\sqrt{\Omega_{\mathrm B}}$. With $Q_1(\cdot,\cdot)$ denoting the first-order Marcum $Q$-function,
\begin{align}
F_{R_{\mathrm A},R_{\mathrm B}}(x,y)
={}&1-e^{-\bar{x}^2}
Q_1(A_\rho\bar{y},B_\rho\bar{x})
\nonumber\\
&-e^{-\bar{y}^2}
\big[1-Q_1(B_\rho\bar{y},A_\rho\bar{x})\big].
\label{eq:biv_rayleigh_cdf}
\end{align}
Consequently, the occupancy probability of the arbitrary rectangle
$\mathcal D_{\rm JC}$ follows by inclusion--exclusion as
\begin{equation}
\begin{aligned}
P_{\rm JC} &= F_{R_{\mathrm A},R_{\mathrm B}}(b,d) - F_{R_{\mathrm A},R_{\mathrm B}}(a,d) - F_{R_{\mathrm A},R_{\mathrm B}}(b,c) \\
&\quad  + F_{R_{\mathrm A},R_{\mathrm B}}(a,c).
\end{aligned}
\label{eq:PDclosed}
\end{equation}

Since a Rayleigh envelope conditioned on the other correlated envelope is Rician, the conditional interval probabilities in~\eqref{eq:lambda_sum} are
\begin{align}
q_{\rm B|A}(x;c,d)
={}&Q_1\!\left(
B_\rho\frac{x}{\sqrt{\Omega_{\mathrm A}}},
A_\rho\frac{c}{\sqrt{\Omega_{\mathrm B}}}
\right)
\nonumber\\
&-Q_1\!\left(
B_\rho\frac{x}{\sqrt{\Omega_{\mathrm A}}},
A_\rho\frac{d}{\sqrt{\Omega_{\mathrm B}}}
\right),
\label{eq:qBA}\\
q_{\rm A|B}(y;a,b)
={}&Q_1\!\left(
B_\rho\frac{y}{\sqrt{\Omega_{\mathrm B}}},
A_\rho\frac{a}{\sqrt{\Omega_{\mathrm A}}}
\right)
\nonumber\\
&-Q_1\!\left(
B_\rho\frac{y}{\sqrt{\Omega_{\mathrm B}}},
A_\rho\frac{b}{\sqrt{\Omega_{\mathrm A}}}
\right).
\label{eq:qAB}
\end{align}

Let
$R_{h_i}(\Delta t)\triangleq
\E[h_i(t)h_i^*(t+\Delta t)]$
denote the channel autocorrelation where $(\cdot)^*$ denotes complex conjugation. Under isotropic/Jakes fading: 
\begin{equation}
R_{h_i}(\Delta t)=\Omega_iJ_0(2\pi f_{m,i}\Delta t),
\qquad i\in\{\mathrm A,\mathrm B\},
\label{eq:jakes}
\end{equation}
where $f_{m,i}$ is the maximum Doppler frequency and $J_0(\cdot)$ is the zeroth-order Bessel function of the first kind. Define the cross-correlation $R_{h_{\rm A}h_{\rm B}}(\Delta t)\triangleq\E[h_{\rm A}(t)h_{\rm B}^*(t+\Delta t)]$, and let the overdot on a correlation function denote differentiation with respect to $\Delta t$. The reduction below assumes $\dot R_{h_i}(0)=\dot R_{h_{\rm A}h_{\rm B}}(0)=0$, i.e., symmetric marginal and cross Doppler spectra, so that, for jointly Gaussian channels, each envelope derivative is independent of the envelope pair. The marginal condition holds for~\eqref{eq:jakes}, and the cross condition must hold for the adopted scattering model. When $f_{m,{\rm A}}\ne f_{m,{\rm B}}$, spectral overlap also caps $\rho$ (e.g., $\rho\le0.56$ for $f_{m,{\rm B}}=0.8f_{m,{\rm A}}$), so the perfectly correlated limit requires equal Doppler spreads.

The standard Rayleigh LCR at amplitude $r$ is \cite{rice1945noise},
\begin{equation}
N_i(r)=
\sqrt{2\pi}\,f_{m,i}\frac{r}{\sqrt{\Omega_i}}
\exp\!\left(-\frac{r^2}{\Omega_i}\right),
\quad i\in\{A,B\}.
\label{eq:ray_lcr}
\end{equation}

Using the derivative independence, $f_{R_{\rm A},R_{\rm B},\dot R_i}=f_{R_{\rm A},R_{\rm B}}f_{\dot R_i}$, where $f_{\dot R_i}$ is the PDF of the envelope derivative $\dot R_i$, each face flux in \eqref{eq:face_lower}--\eqref{eq:face_upper} reduces to the marginal LCR \eqref{eq:ray_lcr} times the probability \eqref{eq:qBA} or \eqref{eq:qAB} that the other envelope lies in its interval, which is the form of \eqref{eq:lambda_sum}. Hence, \eqref{eq:lambda_sum}, \eqref{eq:PDclosed}, and \eqref{eq:jacd_general} give the correlated-Rayleigh JACD for arbitrary node intervals.

\paragraph{Closed-form Rayleigh JAFD}
For this model, $P_{\rm JF}=F_{R_{\rm A},R_{\rm B}}(a,c)$. Define the conditional fade probabilities $\eta_{\rm B|A}(a;c)\triangleq1-Q_1(B_\rho a/\sqrt{\Omega_{\mathrm A}},A_\rho c/\sqrt{\Omega_{\mathrm B}})$ and $\eta_{\rm A|B}(c;a)\triangleq1-Q_1(B_\rho c/\sqrt{\Omega_{\mathrm B}},A_\rho a/\sqrt{\Omega_{\mathrm A}})$.
With the Rayleigh joint fade-entry rate $\Lambda_{\rm JF}(a,c)\triangleq N_{\mathrm A}(a)\eta_{\rm B|A}(a;c)+N_{\mathrm B}(c)\eta_{\rm A|B}(c;a)$,
\begin{equation}
\JAFD(a,c)=\frac{F_{R_{\mathrm A},R_{\mathrm B}}(a,c)}{\Lambda_{\rm JF}(a,c)}.
\label{eq:jafd_rayleigh}
\end{equation}
For identical links with common average power $\Omega$, maximum Doppler frequency $f_m$, and fade threshold $r_{\rm f}$, let $N(\cdot)$, $F(\cdot)$ and $\Gamma_{\rm F}(r_{\rm f})\triangleq F(r_{\rm f})/N(r_{\rm f})$ denote the common Rayleigh LCR, marginal CDF and conventional AFD. At $\rho=0$ the two links provide two independent opportunities for the joint fade to terminate, so $\JAFD(0;r_{\rm f})=\Gamma_{\rm F}(r_{\rm f})/2$; as $\rho\to1$ the trajectories coincide and $\JAFD\to\Gamma_{\rm F}(r_{\rm f})$. The corresponding entry-rate limits, $2N(r_{\rm f})F(r_{\rm f})$ and $N(r_{\rm f})$, are equal when $F(r_{\rm f})=1/2$, i.e., at $r_{\rm f}/\sqrt{\Omega}=\sqrt{\ln 2}$, which explains the correlation-dependent crossing in the numerical results.

\section{Applications and Numerical Evaluation}
Reciprocal SKG is a natural application, since Alice and Bob quantize correlated channel observations into common labels over finite runs. Marginal ACD does not determine how long both observations remain in the same bin, whereas JACD directly characterizes that joint persistence. In the zero-lag model used here, $h_{\rm B}=\sqrt{\rho}\,h_{\rm A}+\sqrt{1-\rho}\,w$, where $w$ is an independent Jakes process with the same maximum Doppler frequency, so both envelopes share the LCR~\eqref{eq:ray_lcr}. This idealizes independent channel-estimation noise low-pass filtered to the Doppler band; with equal in-band signal-to-noise ratio (SNR) at both nodes, $\sqrt{\rho}={\rm SNR}/(1+{\rm SNR})$, so $\rho=0.84$ corresponds to about $10.4$~dB. Other symmetric noise spectra change only the LCRs, through the second spectral moments of the observed processes. A probing delay $\tau$ makes the normalized cross-covariance between $h_{\rm A}$ and $\dot h_{\rm B}$ at most $\sqrt{2}J_1(2\pi f_m\tau)\approx4.4f_m\tau$ (about $0.04$ for $f_m\tau=0.01$), where $J_1(\cdot)$ denotes the first-order Bessel function of the first kind, so the model applies when $f_m\tau\ll1$.

For a common interval $\mathcal I=(r_1,r_2]$, set $a=c=r_1$, $b=d=r_2$, and, in the symmetric case, $\Omega_{\rm A}=\Omega_{\rm B}=\Omega$ and $f_{m,\rm A}=f_{m,\rm B}=f_m$.
By symmetry, the same-bin occupancy is $P_{\rm JC}(\rho;r_1,r_2)=F_{R_{\mathrm A},R_{\mathrm B}}(r_2,r_2)-2F_{R_{\mathrm A},R_{\mathrm B}}(r_1,r_2)+F_{R_{\mathrm A},R_{\mathrm B}}(r_1,r_1)$.
Write $q_{\rho}(x;r_1,r_2)$ for the conditional same-bin probability, i.e. \eqref{eq:qBA} in this symmetric case. The joint-entry rate is then $\Lambda_{\rm JC}(\rho;r_1,r_2)=2[N(r_1)q_\rho(r_1;r_1,r_2)+N(r_2)q_\rho(r_2;r_1,r_2)]$, giving the closed-form symmetric JACD $\JACD(\rho;r_1,r_2)=P_{\rm JC}(\rho;r_1,r_2)/\Lambda_{\rm JC}(\rho;r_1,r_2)$.

For $\rho=0$, let $P_I\triangleq e^{-r_1^2/\Omega}-e^{-r_2^2/\Omega}$ denote the single-node occupancy probability of $\mathcal I$ and $\Xi_{\mathcal I}\triangleq P_I/[N(r_1)+N(r_2)]$ its scalar ACD. Then $q_0(x;r_1,r_2)=P_I$ and $P_{\rm JC}(0;r_1,r_2)=P_I^2$, giving $\JACD(0;r_1,r_2)=\Xi_{\mathcal I}/2$. More generally, for independent nonidentical nodes, $1/\Xi_{\rm JC}=1/\Xi_{\rm AC}+1/\Xi_{\rm BC}$. Under perfect correlation, $\lim_{\rho\to1}\JACD(\rho;r_1,r_2)=\Xi_{\mathcal I}$. 

\subsubsection{JACD-Balanced Multilevel Quantization}
Let $r_{\max}$ be the upper amplitude cutoff and $L$ the number of quantization intervals. Define the quantization thresholds $0=\gamma_0<\gamma_1<\cdots<\gamma_L=r_{\max}$ and $\mathcal I_\ell\triangleq(\gamma_{\ell-1},\gamma_\ell]$, $\ell=1,\ldots,L$. Partitioning the amplitude range by state duration goes back to finite-state Markov channel models~\cite{zhang1999fsmc}. Classical SKG schemes instead use guard-band thresholds with one bit per excursion~\cite{mathur2008radio} or adaptive equiprobable multi-bit quantization~\cite{jana2009effectiveness}. The scalar ACD-based design \cite{9431101} balances the marginal durations $\Xi_{\mathcal I_\ell}$, which does not in general balance the joint durations $\Xi_{{\rm JC},\ell}\triangleq\JACD(\rho;\gamma_{\ell-1},\gamma_\ell)$. We therefore choose the interior thresholds by minimizing the normalized dispersion $\sum_{\ell=1}^{L}(\Xi_{{\rm JC},\ell}/\bar{\Xi}_{\rm JC}-1)^2$, where $\bar{\Xi}_{\rm JC}\triangleq L^{-1}\sum_{\ell=1}^{L}\Xi_{{\rm JC},\ell}$, so that the joint bin durations are approximately balanced. For sampling interval $T_s$, the mean joint sample-run length in bin $\ell$ is approximated at sufficiently fine sampling by $K_{{\rm JC},\ell}\simeq\Xi_{{\rm JC},\ell}/T_s$ \cite{9431101}.

For bin $\ell$, let $P_{{\rm JC},\ell}$ and $\Lambda_{{\rm JC},\ell}$ denote its joint occupancy and entry rate. Since $P_{{\rm JC},\ell}=\Lambda_{{\rm JC},\ell}\Xi_{{\rm JC},\ell}$, equal occupancy does not imply equal persistence: the same probability mass may consist of many short runs or fewer long runs.

\subsubsection{SKG Post-Processing}
Following quantization, information reconciliation (IR) is performed using the blind polar-code protocol of \cite{9184939}: decoding starts at the maximum effective code rate, and each failed attempt discloses a fixed number of additional polar-domain bits until decoding succeeds or the round limit is reached. Failed decoder outputs are not treated as reconciled keys.

\begin{figure}[!t]
\centering
\includegraphics[width=\columnwidth]{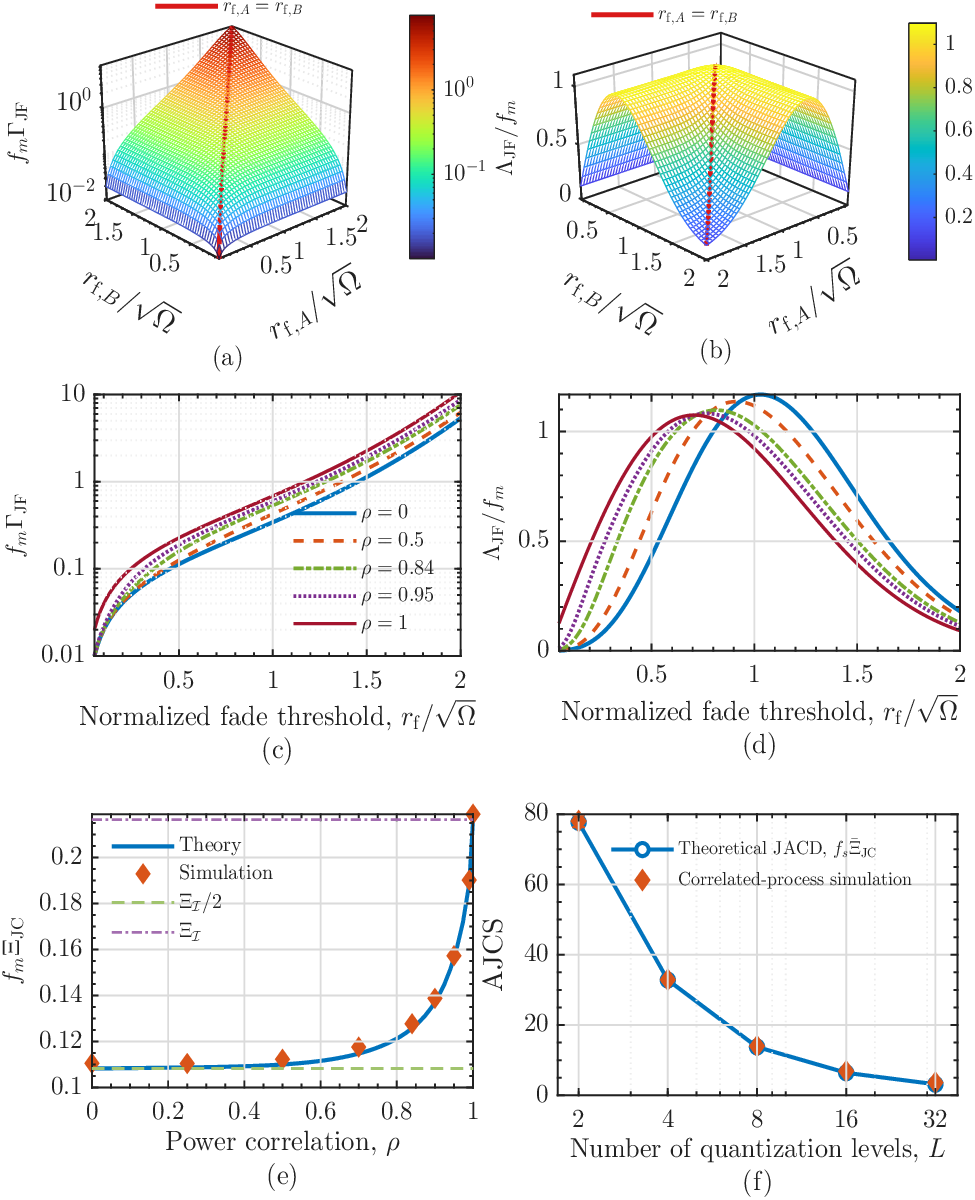}
\caption{Joint-duration characterization and validation. (a) Normalized JAFD and (b) joint fade-entry rate versus unequal fade thresholds at $\rho=0.84$; the diagonal denotes equal thresholds. (c) Normalized JAFD and (d) joint fade-entry rate versus the equal normalized fade threshold for several $\rho$. (e) Analytical and simulated JACD versus $\rho$ for $\mathcal I=(0.5,1]$, with the independent and perfectly correlated limits. (f) Theoretical and simulated average jointly contiguous samples (AJCS) versus the number of JACD-balanced quantization levels.}
\vspace{-12pt}
\label{ref:fig2}
\end{figure}

All amplitudes are normalized by $\sqrt{\Omega}$ and durations by $1/f_m$. Analytical results use \eqref{eq:jacd_general}, \eqref{eq:lambda_sum}, and \eqref{eq:biv_rayleigh_cdf}--\eqref{eq:qAB} for the JACD and \eqref{eq:jafd_rayleigh} for the JAFD. For the analytical JACD validation and joint sample-run study, the sampling frequency $f_s$ is set to $200f_m$.

Figs.~\ref{ref:fig2}(a)--(d) show the normalized JAFD and joint fade-entry rate for unequal thresholds at $\rho=0.84$ and for equal thresholds at several $\rho$. The JAFD increases with the fade threshold, from $\Gamma_{\rm F}(r_{\rm f})/2$ for independent identical links toward the single-link AFD as $\rho\to1$. The entry rate is non-monotonic in the threshold, because enlarging the fade region changes both its occupancy and how often it is entered, and the $\rho=0$ and $\rho\to1$ curves cross at $r_{\rm f}/\sqrt{\Omega}=\sqrt{\ln 2}\simeq0.833$, as predicted. Fig.~\ref{ref:fig2}(e) validates the JACD expression for $\mathcal I=(0.5,1]$ against direct correlated-process simulation. At $\rho=0$, $f_m\Xi_{\rm JC}=0.10824$, exactly one-half of the scalar value $f_m\Xi_{\mathcal I}=0.21648$, and $\Xi_{\rm JC}$ approaches $\Xi_{\mathcal I}$ as $\rho\to1$. The JACD rises by only $3\%$ up to $\rho=0.6$ but by $44\%$ at $\rho=0.95$, so joint persistence changes mostly in the high-correlation regime of reciprocal SKG. In Fig.~\ref{ref:fig2}(f), obtained at $\rho=0.84$ and $r_{\max}=2.5\sqrt{\Omega}$, the bin-averaged JACD $\bar{\Xi}_{\rm JC}$ of JACD-balanced thresholds predicts the average joint sample-run length $\bar K_{\rm JC}\simeq f_s\bar{\Xi}_{\rm JC}$. Direct simulation closely follows this prediction, so the continuous-time JACD accurately estimates the number of consecutive jointly usable samples under fine sampling; conversely, a target run length $\bar K_{\rm JC}$ sets the probing interval $T_s\simeq\bar{\Xi}_{\rm JC}/\bar K_{\rm JC}$.

\begin{figure}[!t]
\centering
\includegraphics[width=\columnwidth]{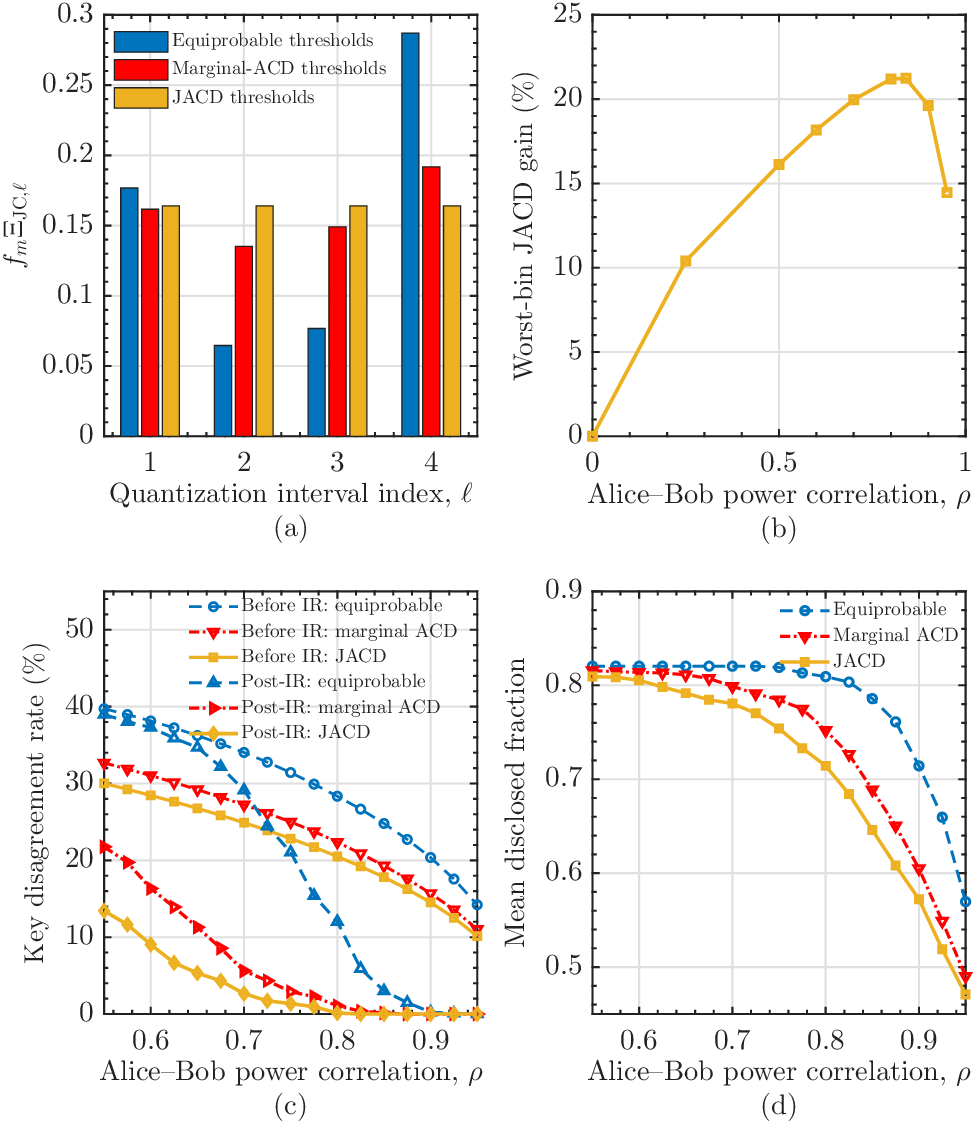}
\caption{JACD-balanced quantization and SKG performance. (a) Normalized same-bin JACDs for equiprobable, marginal-ACD-balanced, and JACD-balanced thresholds at $L=4$, $\rho=0.84$. (b) Worst-bin JACD gain versus Alice--Bob power correlation. For fixed quantizers: (c) raw and residual disagreement after IR~\cite{9184939}, and (d) mean disclosed fraction versus $\rho$.} 
\vspace{-10pt}
\label{ref:fig3}
\end{figure}

For $L=4$, $r_{\max}=2.5\sqrt{\Omega}$, and $\rho=0.84$, Fig.~\ref{ref:fig3}(a) compares equiprobable, marginal-ACD-balanced, and JACD-balanced thresholds, which balance probability mass, single-link persistence, and joint persistence, respectively. JACD balancing equalizes the joint bin durations at $f_m\Xi_{{\rm JC},\ell}\simeq0.16396$, improving the weakest bin by $21.2\%$ over marginal-ACD balancing.

Fig.~\ref{ref:fig3}(b) extends this comparison across Alice--Bob correlation. The gain is zero at $\rho=0$ because $\Xi_{\rm JC}=\Xi_{\mathcal I}/2$, increases as joint and marginal crossing behavior diverge, and decreases again as $\rho\to1$ because JACD approaches the scalar ACD.

For Figs.~\ref{ref:fig3}(c)--(d), we use $L=4$, $r_{\max}=2.5\sqrt{\Omega}$, and $f_s=80f_m$, sweep $\rho$, and keep the three Gray-labeled quantizers fixed, with the JACD thresholds designed at $\rho=0.84$ and the same channel bank for all designs. IR uses polar block length $N=128$, initial effective rate $R_{\max}=0.56$, $\delta=16$ disclosed bits per extra round, and at most three extra rounds. Successful blocks contribute zero residual disagreement, while failed blocks retain their original disagreement. The same reliability ordering, designed from a common mismatch estimate at each $\rho$, serves all three quantizers, so the comparison isolates the effect of the thresholds.

In Fig.~\ref{ref:fig3}(c), the raw disagreement falls as $\rho$ increases and agrees with its closed-form value from~\eqref{eq:biv_rayleigh_cdf} and~\eqref{eq:PDclosed}, e.g., $25.2\%$, $19.6\%$, and $18.0\%$ at $\rho=0.84$ for the three designs. The JACD-balanced quantizer generally yields the smallest residual disagreement, because its lower initial mismatch makes reconciliation more likely to succeed within the permitted rounds. Accordingly, the JACD-balanced quantizer also has the smallest mean disclosed fraction in Fig.~\ref{ref:fig3}(d), i.e., reconciliation typically succeeds with fewer additional disclosures.

\section{Conclusion}
The JACD has been developed for a general $K$-dimensional stationary process. It relates the probability of the joint state to its entry rate and quantifies how long multiple fading envelopes remain within their prescribed amplitude intervals. The JAFD follows as the special case of simultaneous fades. For two correlated Rayleigh envelopes, closed-form expressions were derived using the bivariate Rayleigh CDF, LCRs, and conditional Rician probabilities. The independent and perfectly correlated cases provide exact consistency checks, while the JACD predicts the average length of jointly usable sample runs. Its usefulness was demonstrated in reciprocal SKG, where JACD-balanced thresholds reduced bit disagreement and public disclosure after blind polar codes-based reconciliation. 
\bibliographystyle{IEEEtran}
\bibliography{References.bib}

\end{document}